\documentclass[12pt]{article}

\usepackage{CJKutf8}
\usepackage{amsmath,amssymb,graphicx,float}
\usepackage[unicode]{hyperref}
\usepackage{xcolor}
\usepackage{cite}
\usepackage{indentfirst}

\newcommand{\be}{\begin{equation}\begin{aligned}}
\newcommand{\bea}{\begin{eqnarray}}
\newcommand{\eea}{\end{eqnarray}}
\newcommand{\ba}{\begin{array}}
\newcommand{\ea}{\end{array}}
\newcommand{\ee}{\end{aligned}\end{equation}}

\newcommand{\tr}{\mbox{Tr}}

\begin{document}

\begin{titlepage}

\vspace*{5mm}%

\title{\textbf {The Fused Model of Alternating Spin Chain from
ABJM Theory}}
	\author{Nan Bai$^{a}$\footnote{bainan@mailbox.gxnu.edu.cn}~, Fan Feng$^{a}$\footnote{fanfeng@stu.gxnu.edu.cn}~,  Mao-Zhong Shao$^{a}$\footnote{mzshao@stu.gxnu.edu.cn}~,
}
	\date{}
{\let\newpage\relax\maketitle}
	\maketitle
	\underline{}
	\vspace{-10mm}
	
	\begin{center}
		{\it
            $^{a}$ Department of Physics, Guangxi Normal University, \\Guilin 541004, China
		}
		\vspace{10mm}
	\end{center}

\begin{abstract}
 In this paper we give an algebraic construction of the fused model for ABJM spin chain and find the corresponding boost operator.  We also investigate the open spin Hamiltonian for fused model and point out the general common structures of the boundary terms.
\end{abstract}
\end{titlepage}

\section{Introduction}
ABJM theory is a three-dimensional superconformal Chern-Simons theory with gravity dual as the type IIA superstring theory on $AdS_4\times \mathbb{CP}^3$ background, which is proposed by Aharony, Bergman, Jafferis and Maldacena in 2008 \cite{Aharony:2008ug}. Soon after the discovery, the integrability of ABJM theory has been found by relating the anomalous dimension matrix of the single trace operator made of bi-fundamental fields to a closed alternating spin chain which is shown to be integrable at planar two-loop order in $\rm{SU}(4)$ subsector \cite{Minahan:2008hf,Bak:2008cp}. The integrability was later also extended to the complete $\rm{Osp}(6|4)$ sector and to all-loop order \cite{Zwiebel:2009vb,Minahan:2009te,Gromov:2008qe}.

There are also intensive studies on the integrable models stemming from the original ABJM spin chain with various nontrivial boundaries. For instance, in the orbifold ABJM theory, we will have an integrable closed spin chain with twisted boundary condition \cite{Bai:2016pxs}. In the study of determinant-like operator in ABJM theory, we will treat an open spin chain Hamiltonian \cite{Chen:2018sbp}, whose integrability is  proved by finding a concrete projected K-matrices in the framework of algebraic Bethe ansatz (ABA) \cite{Bai:2019soy}. In the flavored ABJM theory \cite{Hohenegger:2009as,Gaiotto:2009tk,Hikida:2009tp}, we can construct the gauge invariant operator using fundamental/anti-fundamental flavors at two ends without the trace, such operator will also correspond to an open spin chain, which is argued to be integrable by means of coordinate Bethe ansatz (CBA)\cite{Bai:2017jpe}.

In the constructions of integrable models, the so-called boost operator turns out to be an important object which connects different conserved charges through a recursive relation \cite{Tetelman:1982,Sogo:1983,Thacker:1986}. The boost operator can be also used to generate an integrable long range spin chain from a nearest-neighbour spin chain \cite{Bargheer:2008jt}. For the integrable model with regular R-matrix, the boost operator can be easily established \cite{Loebbert:2016cdm,deLeeuw:2021}. However for the non-regular R-matrix, such as two of the four R-matrices adopted in the original ABJM spin chains, the existence of the boost operator is yet unknown. Actually, one of the major motivation for present work is to find a suitable boost operator for ABJM spin chain model.

Motivated by the work \cite{Gombor:2021nhn}, where a general algebraic treatment for medium-range spin chain was proposed, in this paper we will reformulate the original ABJM spin chain model with local three-site interacting Hamiltonian by combining two adjacent quantum spaces into a new single one, and thus obtain the fused ABJM model with nearest-neighbour interactions. We will demonstrate the integrability of the fused model by giving the concrete R-matrix and checking the validity of the Yang-Baxter equation. Due to the regularity of the fused R-matrix, we can obtain the boost operator for the fused model and then use it to analyze the structure of the higher charges. We will also discuss the existence of the boost operators in two sub-chains of the original ABJM model. Finally we will investigate the fused model for open spin chain and try to find out some common structures of the boundary terms by a careful calculation of the open spin chain Hamiltonian.

The paper is organized as follows: In section 2, we briefly review the ABJM spin chain model. In section 3, we present the details of the construction of the fused model for ABJM spin chain and discuss the boost operator in the fused model as well as the original ABJM model. In section 4, we study the fused model for open spin chain and give the concrete spin chain Hamiltonian.  We also analyze the structures of the boundary terms and discuss the integrability of the flavored ABJM spin chain from the algebraic aspects. In the last section, we make the conclusion and point out some future research directions.

\section{Review of ABJM spin chain}
In this section we review the spin chain model originated from ABJM theory. The quantum space in each site of the spin chain is a representation space of $\rm{SU(4)}$ group, alternating from fundamental representation $``\mathbf{4}"$ to anti-fundamental one $``\bar{\mathbf{4}}"$.

There are four kinds of R-matrices ,
\begin{eqnarray}\label{Rmatrix}
\begin{aligned}
R_{ab}(u)=u\mathbb{I}_{ab}+\mathbb{P}_{ab},\quad R_{a\bar{b}}(u)=-(u+2)\mathbb{I}_{a\bar{b}}+\mathbb{K}_{a\bar{b}},\\
R_{\bar{a}\bar{b}}(u)=u\mathbb{I}_{\bar{a}\bar{b}}+\mathbb{P}_{\bar{a}\bar{b}},\quad R_{\bar{a}b}(u)=-(u+2)\mathbb{I}_{\bar{a}b}+\mathbb{K}_{\bar{a}b},
\end{aligned}
\end{eqnarray}
where the subscripts $a$($\bar{a}$) or $b$($\bar{b}$) indicate they belong to $\mathbf{4}$($\bar{\mathbf{4}}$) representation spaces, respectively. $\mathbb{P}$ and $\mathbb{K}$ are permutation and trace operators defined by means of the standard basis matrices $\{e_{ij},i,j=1,2,3,4\}$ as
\begin{eqnarray}\label{PK}
\begin{aligned}
\mathbb{P}=e_{ij}\otimes e_{ji},\quad \mathbb{K}=e_{ij}\otimes e_{ij},
\end{aligned}
\end{eqnarray}
where the repeated indices are summed implicitly. These R-matrices satisfy a total of eight Yang-Baxter equations which are expressed concisely as
\begin{eqnarray}\label{YBE}
\begin{aligned}
R_{AB}(\lambda_A-\lambda_B)R_{AC}(\lambda_A-\lambda_C)R_{BC}(\lambda_B-\lambda_C)=\\
R_{BC}(\lambda_B-\lambda_C)R_{AC}(\lambda_A-\lambda_C)R_{AB}(\lambda_A-\lambda_B),
\end{aligned}
\end{eqnarray}
where $A=\{a,\bar{a}\},\,B=\{b,\bar{b}\},\,C=\{c,\bar{c}\}$.

For the closed alternating spin chain with $2L$ sites, we have the following two monodromy matrices,
\begin{eqnarray}\label{transferc}
\begin{aligned}
T_0(u)&=\tr_{0}R_{01}(u)R_{0\bar{2}}(u)\cdots R_{0,2L-1}(u)R_{0,\overline{2L}}(u),\\
\bar{T}_{\bar{0}}(u)&=\tr_{\bar{0}}R_{\bar{0}1}(u)R_{\bar{0}\bar{2}}(u)\cdots R_{\bar{0},2L-1}(u)R_{\bar{0},\overline{2L}}(u),
\end{aligned}
\end{eqnarray}
where $V_0$ and $V_{\bar{0}}$ are auxiliary spaces and the corresponding transfer matrices are $\tau(u)=\tr_{0}T_{0}(u)$ and $\bar{\tau}(u)=\tr_{\bar{0}}\bar{T}_{\bar{0}}(u)$. Due to the Yang-Baxter relations (\ref{YBE}), the transfer matrices commute with each other for arbitrary spectral parameters,
\begin{eqnarray}
\begin{aligned}
\left[\tau(u),\tau(v)\right]=0,\quad \left[\bar{\tau}(u),\bar{\tau}(v)\right]=0,\quad \left[\tau(u),\bar{\tau}(v)\right]=0.\quad \forall u,v\in \mathbb{C}
\end{aligned}
\end{eqnarray}
The Hamiltonian of the ABJM spin chain model is obtained from $\tau(u)$ and $\bar{\tau}(u)$  as
\begin{eqnarray}\label{abjmH}
\begin{aligned}
H_{\rm{ABJM}}=\frac{d}{du}\log \tau(u)\bigg|_{u=0}+\frac{d}{du}\log \bar{\tau}(u)\bigg|_{u=0},
\end{aligned}
\end{eqnarray}
and a direct computation gives its concrete expression,
\begin{eqnarray}\label{abjmH2}
\begin{aligned}
H_{\rm{ABJM}}=\sum_{l=1}^{2L}\left(\mathbb{P}_{l,l+2}-\frac{1}{2}\mathbb{K}_{l,l+1}\mathbb{P}_{l,l+2}-\frac{1}{2}\mathbb{P}_{l,l+2}\mathbb{K}_{l,l+1}\right),
\end{aligned}
\end{eqnarray}
which, up to an overall prefactor and a constant term, is exactly the anomalous dimension matrix of the dilatational operator in ABJM theory. Furthermore, we see that $H_{\rm{ABJM}}$ is a three-site interacting model with next-to-nearest local Hamiltonian density,
\begin{eqnarray}
\begin{aligned}
h_{l,l+1,l+2}=\mathbb{P}_{l,l+2}-\frac{1}{2}\mathbb{K}_{l,l+1}\mathbb{P}_{l,l+2}-\frac{1}{2}\mathbb{P}_{l,l+2}\mathbb{K}_{l,l+1}.
\end{aligned}
\end{eqnarray}

\section{The construction of fused model}
In this section we present the algebraic construction of the fused model for ABJM spin chain by introducing the Lax operator, the R-matrix and the boost operator.
\subsection{The Lax operator and the R-matrix}

In Eq.\ref{abjmH}, the Hamiltonian of ABJM spin chain is expressed as the sum of the conserved charges from two different transfer matrices. In the following, we will construct a fused model for ABJM spin chain in which the Hamiltonian $H_{\rm{ABJM}}$ is generated from a single transfer matrix. First we multiply the original two monodromy matrices $T_0(u)$ and $\bar{T}_{\bar{0}}(u)$ to get a new one $\mathbf{T}_{0\bar{0}}(u)=T_0(u)\bar{T}_{\bar{0}}(u)$. Then we rearrange the positions of R-matrices in $\mathbf{T}_{0\bar{0}}(u)$ to get
\begin{eqnarray}
\begin{aligned}
\mathbf{T}_{0\bar{0}}(u)=&\left(R_{01}(u)R_{\bar{0}1}(u)R_{0\bar{2}}(u)R_{\bar{0}\bar{2}}(u)\right)\cdots\\
&\left(R_{0,2L-1}(u)R_{\bar{0},2L-1}(u)R_{\bar{2}}(u)R_{\bar{0}\bar{2}}(u)\right).
\end{aligned}
\end{eqnarray}
Thus we find that if we treat the tensor product of two nearest quantum spaces as an enlarged new quantum space, such as  $V_{2j-1,\overline{2j}}=V_{2j-1}\otimes V_{\overline{2j}}$, and also introduce a new auxiliary space as $V_{0\bar{0}}=V_0\otimes V_{\bar{0}}$,  we can define a new Lax operator on the tensor product space $V_{0\bar{0}}\otimes V_{2j-1,\overline{2j}}$ as
\begin{eqnarray}
\begin{aligned}
\mathcal{L}_{(0\bar{0}),(2j-1,\overline{2j})}(u)=R_{0,2j-1}(u)R_{0,\overline{2j}}(u)R_{\bar{0},2j-1}(u)R_{\bar{0},\overline{2j}}(u),
\end{aligned}
\end{eqnarray}
or in more general form,
\begin{eqnarray}
\begin{aligned}
\mathcal{L}_{(a\bar{a}),(b\bar{b})}(u)=R_{ab}(u)R_{a\bar{b}}(u)R_{\bar{a}b}(u)R_{\bar{a}\bar{b}}(u),
\end{aligned}
\end{eqnarray}
 and then $\mathbf{T}_{0\bar{0}}(u)$ becomes
\begin{eqnarray}
\begin{aligned}
\mathbf{T}_{0\bar{0}}(u)=\mathcal{L}_{(0\bar{0}),(1\bar{2})}(u)\mathcal{L}_{(0\bar{0}),(3\bar{4})}(u)\cdots \mathcal{L}_{(0\bar{0}),(2L-1,\overline{2L})}(u).
\end{aligned}
\end{eqnarray}
Hence, we can see that $\mathbf{T}_{0\bar{0}}(u)$ represents a new spin chain of length $L$ with isomorphic auxiliary space and quantum space in each site:
\begin{eqnarray}
\begin{aligned}
V_{0\bar{0}}\cong V_{2j-1,\overline{2j}}=\mathbf{4}\otimes \bar{\mathbf{4}},\quad j=1,2,\cdots,L.
\end{aligned}
\end{eqnarray}
More importantly, the above spin chain is integrable since the following R-matrix
\begin{eqnarray}\label{fuseR}
\begin{aligned}
\mathcal{R}_{(a\bar{a}),(b\bar{b})}(u)=R_{a\bar{b}}(u)R_{ab}(u)R_{\bar{a}\bar{b}}(u)R_{\bar{a}b}(u)
\end{aligned}
\end{eqnarray}
makes the ``RLL" exchange relation holds:
\begin{eqnarray}\label{RLL}
\begin{aligned}
\mathcal{R}_{(a\bar{a}),(b\bar{b})}(u-v)\mathcal{L}_{(a\bar{a}),(c\bar{c})}(u)\mathcal{L}_{(b\bar{b}),(c\bar{c})}(v)=&\\
\mathcal{L}_{(b\bar{b}),(c\bar{c})}(v)\mathcal{L}_{(a\bar{a}),(c\bar{c})}(u)&\mathcal{R}_{(a\bar{a}),(b\bar{b})}(u-v).
\end{aligned}
\end{eqnarray}
As a consistency condition, the R-matrix (\ref{fuseR}) itself should satisfy the Yang-Baxter relation:
\begin{eqnarray}\label{YBE2}
\begin{aligned}
\mathcal{R}_{(a\bar{a}),(b\bar{b})}(u-v)\mathcal{R}_{(a\bar{a}),(c\bar{c})}(u)\mathcal{R}_{(b\bar{b}),(c\bar{c})}(v)=&\\
\mathcal{R}_{(b\bar{b}),(c\bar{c})}(v)\mathcal{R}_{(a\bar{a}),(c\bar{c})}(u)&\mathcal{R}_{(a\bar{a}),(b\bar{b})}(u-v).
\end{aligned}
\end{eqnarray}
Using the Yang-Baxter equations (\ref{YBE}), the above RLL relation (\ref{RLL}) and Yang-Baxter relation (\ref{YBE2}) can be verified  straightforwardly. Notice that the Lax operator $\mathcal{L}_{(a\bar{a}),(b\bar{b})}(u)$, though quite similar to $\mathcal{R}_{(a\bar{a}),(b\bar{b})}(u)$, does not obey the intertwining relation:
\begin{eqnarray}
\begin{aligned}
\mathcal{L}_{(a\bar{a}),(b\bar{b})}(u-v)\mathcal{L}_{(a\bar{a}),(c\bar{c})}(u)\mathcal{L}_{(b\bar{b}),(c\bar{c})}(v)\neq&\\
\mathcal{L}_{(b\bar{b}),(c\bar{c})}(v)\mathcal{L}_{(a\bar{a}),(c\bar{c})}(u)&\mathcal{L}_{(a\bar{a}),(b\bar{b})}(u-v).
\end{aligned}
\end{eqnarray}
We have so far established a new integrable model which will be called the fused model since both the auxiliary space and the quantum space are the fusion of two neighbouring representation spaces. The transfer matrix of the fused model is simply the multiplication of two original ones,
\begin{eqnarray}
\begin{aligned}
t(u)=\tr_{0\bar{0}}\mathbf{T}_{0\bar{0}}(u)=\tau(u)\bar{\tau}(u),
\end{aligned}
\end{eqnarray}
and the Hamiltonian generated from $t(u)$ is just $H_{\rm{ABJM}}$. However, from the viewpoint of the fused model, $H_{\rm{ABJM}}$ turns out to be a nearest-neighbour interacting model
\begin{eqnarray}
\begin{aligned}
H_{\rm{ABJM}}=\sum_{j=1}^L H_{(2j-1,\overline{2j}),(2j+1,\overline{2j+2})},
\end{aligned}
\end{eqnarray}
where the local Hamiltonian density is
\begin{eqnarray}
\begin{aligned}
H_{(2j-1,\overline{2j}),(2j+1,\overline{2j+2})}=h_{2j-1,\overline{2j},2j+1}+h_{\overline{2j},2j+1,\overline{2j+2}}.
\end{aligned}
\end{eqnarray}
Now let us switch back to the original spin chain where the nearest two quantum spaces $\mathbf{4}$ and $\bar{\mathbf{4}}$  are separate sites, then by construction the transfer matrix $t(u)$ for fused model is two-site translation invariant. However we can easily find the Lax operator has the factorized form:
\begin{eqnarray}
\begin{aligned}
\mathcal{L}_{(0\bar{0}),(2j-1,\overline{2j})}(u)=\mathcal{L}_{(0\bar{0}),2j-1}(u)\mathcal{L}_{(0\bar{0}),\overline{2j}}(u),
\end{aligned}
\end{eqnarray}
where
\begin{eqnarray}
\begin{aligned}
\mathcal{L}_{(0\bar{0}),2j-1}(u)&=R_{0,2j-1}(u)R_{\bar{0},2j-1}(u),\\
\mathcal{L}_{(0\bar{0}),\overline{2j}}(u)&=R_{0,\overline{2j}}(u)R_{\bar{0},\overline{2j}}(u),
\end{aligned}
\end{eqnarray}
and thus the transfer matrix can be written as
\begin{eqnarray}\label{facform}
\begin{aligned}
t(u)=\tr_{0\bar{0}}\mathcal{L}_{(0\bar{0}),1}(u)\mathcal{L}_{(0\bar{0}),\bar{2}}(u)\cdots \mathcal{L}_{(0\bar{0}),2L-1}(u)\mathcal{L}_{(0\bar{0}),\overline{2L}}(u),
\end{aligned}
\end{eqnarray}
which is obviously one-site shift invariant.

One last point we would like to mention: We could also use $\mathcal{R}_{(a\bar{a}),(b\bar{b})}(u)$ as the Lax operator to generate a new integrable spin chain, as shown below,
\begin{eqnarray}
\begin{aligned}
\tilde{t}(u)=\tr_{0\bar{0}}\mathcal{R}_{(0\bar{0}),(1\bar{2})}(u)\mathcal{R}_{(0\bar{0}),(3\bar{4})}(u)\cdots \mathcal{R}_{(0\bar{0}),(2L-1,\overline{2L})}(u).
\end{aligned}
\end{eqnarray}
The relation between $t(u)$ and $\tilde{t}(u)$ can be found as follows: Notice that
\begin{eqnarray}
\begin{aligned}
\mathcal{R}_{(0\bar{0}),(2j-1,\overline{2j})}(u)=R_{2j-1,\overline{2j}}(0)\mathcal{L}_{(0\bar{0}),(2j-1,\overline{2j})}(u)R^{-1}_{2j-1,\overline{2j}}(0),
\end{aligned}
\end{eqnarray}
where the similarity transformation matrix $R_{2j-1,\overline{2j}}(0)=-2+\mathbb{K}_{2j-1,\overline{2j}}$ has the properties
\begin{eqnarray}
\begin{aligned}
R_{2j-1,\overline{2j}}(0)^{T}=R_{2j-1,\overline{2j}}(0),\quad R_{2j-1,\overline{2j}}(0)^{T}R_{2j-1,\overline{2j}}(0)=4\mathbb{I}_{2j-1,\overline{2j}},
\end{aligned}
\end{eqnarray}
thus can be seen as a rotation in local quantum space $V_{2j-1,\overline{2j}}=V_{2j-1}\otimes V_{\overline{2j}}$. Then we find $t(u)$ and $\tilde{t}(u)$ are related by a global rotation in the whole Hilbert space $\otimes_{j=1}^L V_{2j-1,\overline{2j}}$,
\begin{eqnarray}
\begin{aligned}
\tilde{t}(u)=\Lambda t(u) \Lambda^{-1},
\end{aligned}
\end{eqnarray}
where
\begin{eqnarray}
\begin{aligned}
\Lambda=R_{1\bar{2}}(0)R_{3\bar{4}}(0)\cdots R_{2L-1,\overline{2L}}(0),
\end{aligned}
\end{eqnarray}
so do the Hamiltonian $\tilde{H}_{\rm{ABJM}}$ obtained from $\tilde{t}(u)$ and $H_{\rm{ABJM}}$: \
\begin{eqnarray}
\begin{aligned}
\tilde{H}_{\rm{ABJM}}=\Lambda H_{\rm{ABJM}} \Lambda^{-1}.
\end{aligned}
\end{eqnarray}
\subsection{Boost operator}
Now we proceed with an investigation of the boost operator for the fused model which can be used to generate higher conserved charges.

From the definition of the fused R-matrix in (\ref{fuseR}), we find
\begin{eqnarray}
\begin{aligned}
\mathcal{R}_{(a\bar{a}),(b\bar{b})}(0)=4\mathbb{P}_{ab}\mathbb{P}_{\bar{a}\bar{b}}.
\end{aligned}
\end{eqnarray}
The R-matrix with the above condition is usually called regular. For the integrable model with regular R-matrix, the way to construct the boost operator is well-known in the literature \cite{Loebbert:2016cdm} and applied to our fused model as follows: First it can be easily shown that the Lax operator is P-symmetric, that is
\begin{eqnarray}
\begin{aligned}
\mathcal{L}_{(a\bar{a}),(b\bar{b})}(u)=\mathcal{L}_{(b\bar{b}),(a\bar{a})}(u),
\end{aligned}
\end{eqnarray}
and thus the RLL relation in (\ref{RLL}) can be written as
\begin{equation}
\begin{aligned}
\mathcal{R}_{(1\bar{2}),(3\bar{4})}(v)\mathcal{L}_{(0\bar{0}),(1\bar{2})}(u+v)\mathcal{L}_{(0\bar{0}),(3\bar{4})}(u)\\
=\mathcal{L}_{(0\bar{0}),(3\bar{4})}(u)\mathcal{L}_{(0\bar{0}),(1\bar{2})}(u+v)\mathcal{R}_{(1\bar{2}),(3\bar{4})}(v).
\end{aligned}
\end{equation}
Then by taking the derivative with respect to $v$ on both sides and setting $v=0$ in the end, we get
\begin{equation}
\begin{aligned}
&\dot{\mathcal{R}}_{(1\bar{2}),(3\bar{4})}(0)\mathcal{L}_{(0\bar{0}),(1\bar{2})}(u)\mathcal{L}_{(0\bar{0}),(3\bar{4})}(u)+
\mathcal{R}_{(1\bar{2}),(3\bar{4})}(0)\dot{\mathcal{L}}_{(0\bar{0}),(1\bar{2})}(u)\mathcal{L}_{(0\bar{0}),(3\bar{4})}(u)\\
=&\mathcal{L}_{(0\bar{0}),(3\bar{4})}(u)\dot{\mathcal{L}}_{(0\bar{0}),(1\bar{2})}(u)\mathcal{R}_{(1\bar{2}),(3\bar{4})}(0)+
\mathcal{L}_{(0\bar{0}),(3\bar{4})}(u)\mathcal{L}_{(0\bar{0}),(1\bar{2})}(u)\dot{\mathcal{R}}_{(1\bar{2}),(3\bar{4})}(0).
\end{aligned}
\end{equation}
Multiplying $\mathcal{R}_{(1\bar{2}),(3\bar{4})}(0)$ on both sides from the left and using the regularity condition $\mathcal{R}_{(1\bar{2}),(3\bar{4})}(0)
=4\mathbb{P}_{13}\mathbb{P}_{\bar{2}\bar{4}}$ gives
\begin{equation}
\begin{aligned}
\left[\frac{1}{16}\mathcal{R}_{(1\bar{2}),(3\bar{4})}(0)\dot{\mathcal{R}}_{(1\bar{2}),(3\bar{4})}(0),\mathcal{L}_{(0\bar{0}),(1\bar{2})}(u)\mathcal{L}_{(0\bar{0}),(3\bar{4})}(u)\right]\\
=-\dot{\mathcal{L}}_{(0\bar{0}),(1\bar{2})}(u)\mathcal{L}_{(0\bar{0}),(3\bar{4})}(u)+\mathcal{L}_{(0\bar{0}),(1\bar{2})}(u)\dot{\mathcal{L}}_{(0\bar{0}),(3\bar{4})}(u),
\end{aligned}
\end{equation}
where $\frac{1}{16}\mathcal{R}_{(1\bar{2}),(3\bar{4})}(0)\dot{\mathcal{R}}_{1\bar{2},3\bar{4}}(0)$ is just the local Hamiltonian $H_{(1\bar{2}),(3\bar{4})}$ plus an identity operator and thus can be replaced by $H_{(1\bar{2}),(3\bar{4})}$ in the commutation relation. By the substitution of the indices:
$1\rightarrow 2k-1,\, 2\rightarrow 2k,\, 3\rightarrow 2k+1,\, 4\rightarrow 2k+2$, we have
\begin{equation}
\begin{aligned}
&\left[{H}_{(2k-1,\overline{2k}),(2k+1,\overline{2k+2})},\mathcal{L}_{(0\bar{0}),(2k-1,\overline{2k})}(u)\mathcal{L}_{(0\bar{0}),(2k+1,\overline{2k+2})}(u)\right]\\
=&-\dot{\mathcal{L}}_{(0\bar{0}),{(2k-1,\overline{2k})}}(u)\mathcal{L}_{(0\bar{0}),{(2k+1,\overline{2k+2})}}(u)+\mathcal{L}_{(0\bar{0}),(2k-1,\overline{2k})}(u)
\dot{\mathcal{L}}_{(0\bar{0}),(2k+1,\overline{2k+2})}(u).
\end{aligned}
\end{equation}
Then, by multiplying $\prod _{j=1}^{k-1}\mathcal{L}_{(0\bar{0}),(2j-1,\overline{2j})}(u)$ from the left and  $\prod _{j=k+2}^{L}\mathcal{L}_{(0\bar{0}),(2j-1,\overline{2j})}(u)$ from the right to both sides of the above equation, we get
\begin{equation}
\begin{aligned}
&\left[{H}_{(2k-1,\overline{2k}),(2k+1,\overline{2k+2})},\mathbf{T}_{0\bar{0}}(u)\right]\\
=&-\left[\prod _{j=1}^{k-1}\mathcal{L}_{(0\bar{0}),(2j-1,\overline{2j})}(u)\right]\dot{\mathcal{L}}_{(0\bar{0}),{(2k-1,\overline{2k})}}(u)
\left[\prod _{j=k+1}^{L}\mathcal{L}_{(0\bar{0}),(2j-1,\overline{2j})}(u)\right]\\
&+\left[\prod _{j=1}^{k}\mathcal{L}_{(0\bar{0}),(2j-1,\overline{2j})}(u)\right]\dot{\mathcal{L}}_{(0\bar{0}),{(2k+1,\overline{2k+2})}}(u)
\left[\prod _{j=k+2}^{L}\mathcal{L}_{(0\bar{0}),(2j-1,\overline{2j})}(u)\right].
\end{aligned}
\end{equation}
Finally, by summing up the above equation for each $k=1,2,\cdots, L-1$, we find
\begin{equation}
\begin{aligned}
&\left[\sum_{k=1}^{L-1}k{H}_{(2k-1,\overline{2k}),(2k+1,\overline{2k+2})},{\mathbf{T}}_{0\bar{0}}(u)\right]\\
=&-\frac{d{\mathbf{T}}_{0\bar{0}}(u)}{du}+L\left[\prod _{j=1}^{L-1}\mathcal{L}_{0\bar{0},(2j-1,\overline{2j})}(u)\right]\dot{\mathcal{L}}_{0\bar{0},(2L-1,\overline{2L})}(u).
\end{aligned}
\end{equation}
For an infinite spin chain or closed spin chain, we get
\begin{equation}\label{boost1}
\begin{aligned}
\frac{dt(u)}{du}=\left[\mathcal{B},t(u)\right],
\end{aligned}
\end{equation}
where $\mathcal{B}$ is the boost operator defined as
\begin{equation}
\begin{aligned}
\mathcal{B}=-\sum_k k {H}_{(2k-1,\overline{2k}),(2k+1,\overline{2k+2})}.
\end{aligned}
\end{equation}
The conserved charges are defined as the coefficients of the Taylor expansion of $\log t(u)$ at $u=0$,
\begin{equation}
\begin{aligned}
\log t(u)=i\sum_{n=0}Q_{n+1}u^n,
\end{aligned}
\end{equation}
then the first charge $Q_1$ is simply $-i\log t(0)$ which is non-local, since
\begin{equation}
\begin{aligned}
t(0)=4^L\mathbb{P}_{2L-3,2L-1}\cdots \mathbb{P}_{13}\mathbb{P}_{\overline{2L-2},\overline{2L}}\cdots \mathbb{P}_{\bar{2}\bar{4}}
\end{aligned}
\end{equation}
is the shift operator acting on the whole spin chain. The second charge $Q_2$ corresponds to the spin chain Hamiltonian:
\begin{equation}
\begin{aligned}
iQ_2=\frac{d}{du}\log t(u)\bigg|_{u=0}=H_{\rm{ABJM}}.
\end{aligned}
\end{equation}
The rest of the higher charges can be derived by the boost operator $\mathcal{B}$ from the relation (\ref{boost1}) as:
\begin{equation}
\begin{aligned}
Q_{n+1}=\frac{1}{n}\left[\mathcal{B},Q_n\right],\quad n=2,3,\cdots
\end{aligned}
\end{equation}
thus the next charge $Q_3$ is found to be
\begin{equation}
\begin{aligned}
2Q_3&=-i\sum_j\left[H_{(2j-1,\overline{2j}),(2j+1,\overline{2j+2})},\,\,H_{(2j+1,\overline{2j+2}),(2j+3,\overline{2j+4})}\right]\\
&=-i\sum_j\left\{\left[h_{\overline{2j},2j+1,\overline{2j+2}},\,\,h_{2j+1,\overline{2j+2},2j+3}+h_{\overline{2j+2},2j+3,\overline{2j+4}}\right]+\Delta(j)\right\},
\end{aligned}
\end{equation}
where $\Delta(j)$ is a local operator with interaction range over five sites:
\begin{equation}
\begin{aligned}
\Delta(j)=\left[h_{2j-1,\overline{2j},2j+1},\,h_{2j+1,\overline{2j+2},2j+3}\right].
\end{aligned}
\end{equation}
The form of the third charge $Q_3$ shown above seemingly violates the generalized Reshetikhin condition proposed in \cite{Gombor:2021nhn} for integrable three-site model, in which $\Delta(j)$ is conjectured to be a three-site operator. However, we would like the emphasize that our fused model is essentially a two-site model with auxiliary space and quantum space isomorphic to the tensor product space $\mathbf{4}\otimes \bar{\mathbf{4}}$, and thus even in its factorized form (\ref{facform}), we need two different reduced Lax operators $\mathcal{L}_{(a\bar{a}),b}(u)$ and $\mathcal{L}_{(a\bar{a}),\bar{b}}(u)$, which are not regular either,
\begin{equation}
\begin{aligned}
\mathcal{L}_{(a\bar{a}),b}(0)=\mathbb{P}_{ab}(-2+\mathbb{K}_{\bar{a}b}),\quad \mathcal{L}_{(a\bar{a}),\bar{b}}(0)=(-2+\mathbb{K}_{a\bar{b}})\mathbb{P}_{\bar{a}\bar{b}}.
\end{aligned}
\end{equation}
These major differences indicate the fused model investigated in this work does not belong to the normal three-site interacting model considered in \cite{Gombor:2021nhn}, and thus does not obey the conjecture.

As a final remark on the boost operator of ABJM spin chain, let us consider the possibility of the existence of boost operator in either of the two sub-chains, $\tau(u)$ or $\bar{\tau}(u)$. Suppose the boost operator we are looking for is a ``matrix" type operator and is composed of the local inhomogeneous density, then we can write it in a very general form, for instance, the boost operator $\mathfrak{b}$ for $\tau(u)$ can be expressed as,
\begin{equation}
\begin{aligned}
\mathfrak{b}=\sum_j f(j) b_{j,j+1,\cdots,j+l-1},
\end{aligned}
\end{equation}
where $f(j)$ is a function of the site position representing the inhomogeneity of the operator and $b_{j,j+1,\cdots,j+l-1}$ is the local density with the interaction range over $l$ sites starting at $j$-th site. The boost operator $\mathfrak{b}$ should satisfy the condition:
\begin{equation}
\begin{aligned}
\frac{d\tau(u)}{du}=\left[\mathfrak{b},\tau(u)\right],\quad \forall u\in \mathbb{C}.
\end{aligned}
\end{equation}
At special point $u=0$, since $\left[\tau^{-1}(0),\dot{\tau}(0)\right]=0$, the above equation will reduce to
\begin{equation}\label{range}
\begin{aligned}
\tau^{-1}(0)\mathfrak{b}\tau(0)+\tau(0)\mathfrak{b}\tau^{-1}(0)=2\mathfrak{b}.
\end{aligned}
\end{equation}
Note that $\tau(0)$ is no longer a shift operator but has the following form
\begin{equation}
\begin{aligned}
\tau(0)=\left(-2+\mathbb{K}_{1\bar{2}}\right)\cdots \left(-2+\mathbb{K}_{2L-1,\overline{2L}}\right)\mathbb{P}_{2L-3,2L-1}\cdots \mathbb{P}_{35}\mathbb{P}_{13},
\end{aligned}
\end{equation}
then it follows that $\tau(0) b_{j,j+1,\cdots,j+l-1}\tau^{-1}(0)$ not only shifts the sites that  $b_{j,j+1,\cdots,j+l-1}$ acts, but also increases its interaction range. As a result, the interaction range of the local density on both sides of (\ref{range}) cannot be equal. So we conclude that there is no matrix type boost operator with inhomogeneous local density for integrable spin chain $\tau(u)$ or $\bar{\tau}(u)$.

\section{Open spin chain Hamiltonian from fused model}
In this section, we turn to the study of open spin chain Hamiltonian for fused model. First, let us review the construction of the Hamiltonian for ordinary $2L$-sites alternating spin chain with open boundaries. We need the following two so-called double row transfer matrices:
\begin{equation}
\begin{aligned}
\tau_b(u)=\tr_0 K^{+}_{0}(u)T_0(u)K^-_0(u)T^{-1}_0(-u),\\
\bar{\tau}_b(u)=\tr_{\bar{0}}\bar{K}^+_{\bar{0}}(u)\bar{T}_{\bar{0}}(u)\bar{K}_{\bar{0}}^-(u)\bar{T}^{-1}_{\bar{0}}(-u),
\end{aligned}
\end{equation}
where the symbol ``b" is used to distinguish them from the closed spin chain transfer matrices defined in \ref{transferc}. $K_0^+(u)(\bar{K}^+_{\bar{0}}(u))$ and $K^-_0(u)(\bar{K}^-_{\bar{0}}(u))$ are reflection matrices accounting for the left and right boundary local Hamiltonian, respectively. The two right boundary reflection matrices $K^-_0(u)$ and $\bar{K}^-_{\bar{0}}(u)$ satisfy four reflection equations (REs) given below:
\begin{equation}
\begin{aligned}
R_{ab}(u-v)K^-_a(u)R_{ba}(u+v)K^-_b(v)=K^-_b(v)R_{ab}(u+v)K^-_a(u)R_{ba}(u-v),\\
R_{\bar{a}\bar{b}}(u-v)\bar{K}^-_{\bar{a}}(u)R_{\bar{b}\bar{a}}(u+v)\bar{K}^-_{\bar{b}}(v)=\bar{K}^-_{\bar{b}}(v)R_{\bar{a}\bar{b}}(u+v)\bar{K}^-_{\bar{a}}(u)R_{\bar{b}\bar{a}}(u-v),\\
R_{a\bar{b}}(u-v)K^-_a(u)R_{\bar{b}a}(u+v)\bar{K}^-_{\bar{b}}(v)=\bar{K}^-_{\bar{b}}(v)R_{a\bar{b}}(u+v)K^-_a(u)R_{\bar{b}a}(u-v),\\
R_{\bar{a}b}(u-v)\bar{K}^-_{\bar{a}}(u)R_{b\bar{a}}(u+v)K^-_b(v)=K^-_b(v)R_{\bar{a}b}(u+v)\bar{K}^-_{\bar{a}}(u)R_{b\bar{a}}(u-v),
\end{aligned}
\end{equation}
while the other two left boundary reflection matrices $K^+_0(u)$ and $\bar{K}^+_{\bar{0}}(u)$ satisfy similar dual reflection equations and can be obtained from $K^-_0(u)$ and $\bar{K}^-_{\bar{0}}(u)$ by some isomorphism transformations. Due to these reflection relations, the transfer matrices form the commutating class,
\begin{equation}
\begin{aligned}
\left[\tau_b(u),\tau_b(v)\right]=0,\quad \left[\bar{\tau}_b(u),\bar{\tau}_b(v)\right]=0,\quad \left[\tau_b(u),\bar{\tau}_b(v)\right]=0,\quad \forall u,v\in \mathbb{C}
\end{aligned}
\end{equation}
showing the integrability of the open spin chain model. Then the boundary Hamiltonian from the original two open spin chains is given as
\begin{equation}\label{bm1}
\begin{aligned}
H_{b}=\frac{d}{du}\log \left(\tau_b(u)\bar{\tau}_b(u)\right)\bigg|_{u=0}.
\end{aligned}
\end{equation}
Now we consider the open spin chain from fused model. Since the quantum space and auxiliary space are both $\mathbf{4}\otimes\bar{\mathbf{4}}$, we thus introduce two reflection matrices $K^+_{0\bar{0}}(u)$ and $K^-_{0\bar{0}}(u)$ defined on tensor product space $V_0\otimes V_{\bar{0}}$, of which $K^-_{0\bar{0}}(u)$ satisfy the reflection equation,
\begin{equation}
\begin{aligned}
\mathcal{R}_{(a\bar{a}),(b\bar{b})}(u-v)K^-_{a\bar{a}}(u)\mathcal{R}_{(b\bar{b}),(a\bar{a})}(u+v)K^-_{b\bar{b}}(v)\\
=K^-_{b\bar{b}}(v)\mathcal{R}_{(a\bar{a}),(b\bar{b})}(u+v)K^-_{a\bar{a}}(u)\mathcal{R}_{(b\bar{b}),(a\bar{a})}(u-v),
\end{aligned}
\end{equation}
and $K^+_{0\bar{0}}(u)$ satisfy similar dual reflection equation. The double row transfer matrix for our fused model is
\begin{equation}
\begin{aligned}
t_b(u)=\tr_{0\bar{0}}K^{+}_{0\bar{0}}(u)\mathbf{T}_{0\bar{0}}(u)K^-_{0\bar{0}}(u)\mathbf{T}^{-1}_{0\bar{0}}(-u),
\end{aligned}
\end{equation}
and the open spin chain Hamiltonian is then obtained as
\begin{equation}\label{Ham}
\begin{aligned}
\tilde{H}_b=\frac{d}{du}\log t_b(u)\bigg|_{u=0}.
\end{aligned}
\end{equation}
We can see that, unlike the closed spin chain case, the fused Hamiltonian $\tilde{H}_b$ is different from the Hamiltonian $H_b$ composed of  two original spin chains, hence representing two different open spin chain models. Though in very rare cases, with highly constrained reflection matrices, the two open spin chain models could be equivalent. For instance, if we assume that $K^+_{0\bar{0}}(u)=K^+_0(u)\bar{K}^+_{\bar{0}}(u)$ and $K^-_{0\bar{0}}(u)=K^-_0(u)\bar{K}^-_{\bar{0}}(u)$, then the transfer matrix reduces to:
\begin{equation}
\begin{aligned}
t_b(u)&=\tr_{0\bar{0}} K^+_0(u)\bar{K}^+_{\bar{0}}(u)T_0(u)\bar{T}_{\bar{0}}(u)K^-_0(u)\bar{K}^-_{\bar{0}}(u)\bar{T}_{\bar{0}}^{-1}(-u)T^{-1}_0(-u)\\
&=\tr_{0\bar{0}}K^+_0(u)T_0(u)K^-_0(u)\left[\bar{K}^+_{\bar{0}}(u)\bar{T}_{\bar{0}}(u)\bar{K}^-_{\bar{0}}(u)\bar{T}_{\bar{0}}^{-1}(-u)\right]T^{-1}_0(-u).
\end{aligned}
\end{equation}
Therefore, if we further impose the following condition
\begin{equation}
\begin{aligned}
\left[\bar{K}^+_{\bar{0}}(u)\bar{T}_{\bar{0}}(u)\bar{K}^-_{\bar{0}}(u)\bar{T}_{\bar{0}}^{-1}(-u),\,T^{-1}_0(-u)\right]=0,
\end{aligned}
\end{equation}
then we will have $t_b(u)=\tau_b(u)\bar{\tau}_b(u)$, which leads to the same Hamiltonian.

We also note that, in order to deal with the open spin chain model with degrees of freedom on the boundary, the reflection K-matrices will also act on an additional internal space. By tracing over the auxiliary spaces $V_0\otimes V_{\bar{0}}$, the interactions between the boundary and bulk can be achieved. Hence, in the most general settings, the reflection matrices should be treated as an operator-valued matrix on auxiliary space.

Finally, let us discuss the concrete open spin chain Hamiltonian derived from the fused model transfer matrix (\ref{Ham}). Since the whole calculation is straightforward but quite tedious, here we directly give the final results. For an open spin chain of length $2L$ with internal degrees of freedom acting on the boundaries,  the complete Hamiltonian $\tilde{H}_b$ consists of three parts: the left boundary term $H_l$, the bulk Hamiltonian $H_{\rm{in}}$ and the right boundary term $H_r$,
\begin{equation}
\begin{aligned}
\tilde{H}_b=H_{l}+H_{\rm{in}}+H_r.
\end{aligned}
\end{equation}
The bulk part is just the ordinary closed spin chain Hamiltonian
\begin{equation}
\begin{aligned}
H_{\rm{in}}=\sum_{l=1}^{2L-4}\left(2\mathbb{P}_{l,l+2}-\mathbb{P}_{l,l+2}\mathbb{K}_{l,l+1}-\mathbb{K}_{l,l+1}\mathbb{P}_{l,l+2}\right).
\end{aligned}
\end{equation}
The left boundary term can be further organized into three parts:
\begin{equation}
\begin{aligned}
H_l=H_{l1}+H_{l2}+H_{l3}.
\end{aligned}
\end{equation}
Among them, $H_{l1}$ is a pure left boundary term, which acts trivially on the bulk Hilbert space, given as
\begin{equation}
\begin{aligned}
H_{l1}=&\left[\tr_{0\bar{0}}K^+_{0\bar{0}}(0)\right]^{-1}\left[\tr_{0\bar{0}}\frac{dK^+_{0\bar{0}}(u)}{du}\bigg|_{u=0}\right]\\
-&\frac{1}{2}\left[\tr_{0\bar{0}}K^+_{0\bar{0}}(0)\right]^{-1}\left[\tr_{0\bar{0}}K^+_{0\bar{0}}(0)\left(-2+\mathbb{K}_{0\bar{0}}\right)\right];
\end{aligned}
\end{equation}
$H_{l2}$ is in fact an bulk term acting on the leftmost two sites,
\begin{equation}
\begin{aligned}
H_{l2}=-\frac{1}{2}\left(-2+\mathbb{K}_{12}\right);
\end{aligned}
\end{equation}
$H_{l3}$, representing the true bulk-boundary interaction, has the following form:
\begin{equation}
\begin{aligned}
H_{l3}=&2\left[\tr_{0\bar{0}}K^+_{0\bar{0}}(0)\right]^{-1}\left[\tr_{0\bar{0}}K^+_{0\bar{0}}(0)\mathbb{P}_{01}\right]\\
+&\frac{1}{8}\left[\tr_{0\bar{0}}K^+_{0\bar{0}}(0)\right]^{-1}\left(-2+\mathbb{K}_{1\bar{2}}\right)
\left[\tr_{0\bar{0}}K^+_{0\bar{0}}(0)\left(-2+\mathbb{K}_{0\bar{0}}\right)\mathbb{P}_{\bar{0}\bar{2}}\left(-2+\mathbb{K}_{0\bar{0}}\right)\right]\\
\times &\left(-2+\mathbb{K}_{1\bar{2}}\right).
\end{aligned}
\end{equation}
The right boundary term can be divided into two parts,
\begin{equation}
\begin{aligned}
H_{r}=H_{r1}+H_{r2},
\end{aligned}
\end{equation}
where $H_{r1}$ can be seen as the remaining part of the bulk Hamiltonian acting on the rightmost several sites, shown below,
\begin{equation}
\begin{aligned}
H_{r1}= &\,
\mathbb{P}_{\overline{2L-2},\overline{2L}}-\frac{1}{2}\mathbb{P}_{\overline{2L-2},\overline{2L}}\mathbb{K}_{2L-1,\overline{2L}}-
\frac{1}{2}\mathbb{K}_{2L-1,\overline{2L}}\mathbb{P}_{\overline{2L-2},\overline{2L}}\\
+&\mathbb{P}_{2L-3,2L-1}-\frac{1}{2}\mathbb{P}_{2L-3,2L-1}\mathbb{K}_{2L-3,\overline{2L-2}}-\frac{1}{2}\mathbb{K}_{2L-3,\overline{2L-2}}\mathbb{P}_{2L-3,2L-1}\\
-&\frac{1}{4}\mathbb{K}_{2L-3,\overline{2L-2}}.
\end{aligned}
\end{equation}
$H_{r2}$ includes the interaction between the bulk and the right boundary internal degrees of freedom. Since the expression of $H_{r2}$ is quite lengthy, let us first define the following quantity,
\begin{equation}
\begin{aligned}
\Delta=\left(-2+\mathbb{K}_{2L-1,\overline{2L}}\right)K^-_{2L-1,\overline{2L}}(0)\left(-2+\mathbb{K}_{2L-1,\overline{2L}}\right),
\end{aligned}
\end{equation}
then $H_{r2}$ can be expressed in terms of following four parts,
\begin{equation}
\begin{aligned}
H_{r2}=\Delta_1+\Delta_2+\Delta_3+\Delta_4,
\end{aligned}
\end{equation}
where
\begin{equation}
\begin{aligned}
\Delta_1=&-\frac{1}{4}\Delta^{-1}\left(-2+\mathbb{K}_{2L-1,\overline{2L}}\right)\Delta ,\\
\Delta_2=&\left(-2+\mathbb{K}_{2L-1,\overline{2L}}\right)^{-1}\left[K^-_{2L-1,\overline{2L}}(0)\right]^{-1}\left[\frac{dK^-_{2L-1,\overline{2L}}(0)}{du}\bigg|_{u=0}\right]\\
&\times\left(-2+\mathbb{K}_{2L-1,\overline{2L}}\right),\\
\Delta_3=&\Delta^{-1}\left(\mathbb{P}_{2L-3,2L-1}-\frac{1}{2}\mathbb{P}_{2L-3,2L-1}\mathbb{K}_{2L-3,\overline{2L-2}}-\frac{1}{2}\mathbb{K}_{2L-3,\overline{2L-2}}\mathbb{P}_{2L-3,2L-1}
\right.\\&\left.+\frac{1}{4}\mathbb{K}_{2L-3,\overline{2L-2}}\right)\Delta,\\
\Delta_4=&\Delta^{-1}\left(\mathbb{P}_{\overline{2L-2},\overline{2L}}-\frac{1}{2}\mathbb{P}_{\overline{2L-2},\overline{2L}}\mathbb{K}_{2L-1,\overline{2L}}-
\frac{1}{2}\mathbb{K}_{2L-1,\overline{2L}}\mathbb{P}_{\overline{2L-2},\overline{2L}}\right.\\
&\left.+\frac{1}{4}\mathbb{K}_{2L-1,\overline{2L}}\right)\Delta.
\end{aligned}
\end{equation}
As shown above, the boundary Hamiltonians have very complicated forms, and by choosing different reflection K-matrices, we will have various boundary terms. However, we can find some common structures by analyzing the indices of the components of the boundary Hamiltonian. To be concrete, let us focus on the left boundary, especially the nontrivial bulk-boundary interaction term $H_{l3}$, which acts  on the boundary internal space $V_{\rm{in}}$ and two leftmost quantum spaces $V_1$ and $V_{\bar{2}}$,
\begin{equation}
\begin{aligned}
H_{l3}\in \rm{End}\left(V_{in}\otimes V_1\otimes V_{\bar{2}}\right).
\end{aligned}
\end{equation}
Note that the internal space $V_{\rm{in}}$ needn't to be isomorphic to $V_1$ or $V_{\bar{2}}$ and thus can have different dimension.
Now we look at the component of $H_{l3}$:
$\left[H_{l3}\right]_{i,J_1,I_2}^{j,I_1,J_2}$,
where $\{i,j\}\in V_{\rm{in}},\,\{I_1,J_1\}\in V_1,\{I_2,J_2\}\in V_{\bar{2}}$. There are several terms in $H_{l3}$ and we will discuss them separately.
First, for notational convenience, let us write
\begin{equation}
\begin{aligned}
B&=\left[\tr_{0\bar{0}}K^+_{0\bar{0}}(0)\right]^{-1}\in \rm{End} (V_{in}),\\
M&=\tr_{0\bar{0}}K^+_{0\bar{0}}(0)\mathbb{P}_{01}  \in\rm{End} (V_{in}\otimes V_1),\\
S&=\tr_{0\bar{0}}K^+_{0\bar{0}}(0)\left(-2+\mathbb{K}_{0\bar{0}}\right)\mathbb{P}_{\bar{0}\bar{2}}\left(-2+\mathbb{K}_{0\bar{0}}\right) \in\rm{End} (V_{in}\otimes V_{\bar{2}}),
\end{aligned}
\end{equation}
then the component of each term in $H_{l3}$ reads:
\begin{equation}\label{lb1}
\begin{aligned}
\left(BM\right)_{i,j_1,I_2}^{j,I_1,J_2}&=\left(BM\right)_{i,j_1}^{j,I_1}\cdot \delta_{I_2}^{J_2},\\
\left(BS\right)_{i,j_1,i_2}^{j,i_1,J_2}&=\left(BS\right)_{i,I_2}^{j,J_2}\cdot \delta^{I_1}_{J_1},\\
\left(B\mathbb{K}_{1\bar{2}}S\right)_{i,J_1,I_2}^{j,I_1,J_2}&=\left(B_i^k S_{k,J_1}^{j,J_2}\right)\cdot \delta^{I_1}_{I_2},\\
\left(BS\mathbb{K}_{1\bar{2}}\right)_{i,J_1,I_2}^{j,I_1,J_2}&=\left(B_i^k S_{k,I_2}^{j,I_1}\right)\cdot \delta^{J_2}_{J_1},\\
\left(B\mathbb{K}_{1\bar{2}}S\mathbb{K}_{1\bar{2}}\right)_{i,J_1,I_2}^{j,I_1,J_2}&= \left(B_i^k S_{k,L}^{j,L}\right)\cdot \delta^{I_1}_{I_2}\delta^{J_2}_{J_1},
\end{aligned}
\end{equation}
where we have used the component forms of $\mathbb{K}$: $\mathbb{K}_{J_1,I_2}^{I_1,J_2}=\delta^{I_1}_{I_2}\delta_{J_1}^{J_2}$. For the other two left boundary terms, we can easily find
\begin{equation}\label{lb2}
\begin{aligned}
\left(H_{l1}\right)_{i,J_1,I_2}^{j,I_1,J_2}&=\left(H_{l1}\right)_i^j\cdot \delta^{I_1}_{J_1}\delta_{I_2}^{J_2},\\
\left(H_{l2}\right)_{i,J_1,I_2}^{j,I_1,J_2}&=\left(H_{l2}\right)^{I_1,J_2}_{J_1,I_2}\cdot \delta_i^j.
\end{aligned}
\end{equation}
Thus we find, for each of the left boundary terms, there exists a unique universal Kronecker delta factor, independent of the specific selection of the K-matrix. In other words, given an open spin chain Hamiltonian, if the components of the left boundary terms do not belong to the boundary types shown above, then such an open spin chain cannot be an integrable spin chain, at least not one originated from our fused model.

For a concrete example, we may consider the open spin chain Hamiltonian from flavored ABJM theory \cite{Bai:2017jpe}. Due to the coupling between the bulk bi-fundamental fields and the boundary fundamental flavors, the bulk $SU(4)$ R-symmetry will break into a remaining $SU(2)_R$ and a diagonal subgroup $SU(2)_D$:
\begin{equation}
\begin{aligned}
SU(4)_R \rightarrow  SU(2)_R\times SU(2)_D
\end{aligned}
\end{equation}
In this case, the boundary internal space $V_{\rm{in}}$ is the two dimensional fundamental representation space of $SU(2)_R$ , i.e. $V_{\rm{in}}=\mathbf{2}\cong \bar{\mathbf{2}}$, $i,j\in\left\{1,2\right\}$; while $V_{1}$ and $V_{\bar{2}}$ are the four-dimensional representation spaces $\mathbf{2}\times \mathbf{2}$ and $\bar{\mathbf{2}}\times \bar{\mathbf{2}}$ of $SU(2)_R\times SU(2)_D$, respectively, whose component indices can be formulated by a pair of $SU(2)$ indices, i.e. $I=iA,J=jB, \,i,j\in\{1,2\},\,A,B\in\{1,2\}$, and thus the delta function is simply given as:
$\delta^I_J=\delta^i_j\delta^A_B$. Then it is a simple task to rewrite the boundary terms in (\ref{lb1}) and (\ref{lb2}) using the composite $SU(2)_R\times SU(2)_D$ indices to replace $SU(4)_R$ indices, i.e. $I\rightarrow(iA)$. The Hamiltonian of flavored ABJM spin chain has the following three types of left boundary terms \cite{Bai:2017jpe}:
\begin{equation}
\begin{aligned}
{\rm{type}}\,1: \qquad\quad\quad \delta^{A_1}_{A_2}\delta^{B_2}_{B_1}\delta^{j_2}_{i}\delta^{i_1}_{j_1}\delta^j_{i_2},\\
{\rm{type}}\,2: \qquad\quad\quad \delta^{A_1}_{B_1}\delta^{B_2}_{A_2}\delta^{j_2}_{i}\delta^{i_1}_{j_1}\delta^j_{i_2},\\
{\rm{type}}\,3: \qquad\quad\quad \delta^{A_1}_{B_1}\delta^{B_2}_{A_2}\delta^{i_1}_{i}\delta^{j}_{j_1}\delta^{j_2}_{i_2}.
\end{aligned}
\end{equation}
We can easily observe that the type 2 and type 3 terms have the factors $\delta^{I_{1}}_{J_1}=\delta^{A_1}_{B_1}\delta^{i_1}_{j_1}$ and
$\delta_{I_2}^{J_2}=\delta^{B_2}_{A_2}\delta^{j_2}_{i_2}$, respectively, thus could be obtained from $BM$ and $BS$ terms in (\ref{lb1}). As for type 1 term,  it mixes the indices of all three spaces $V_{\rm{in}}\otimes V_1\otimes V_{\bar{2}}$ and does not belong to any type of boundary terms in (\ref{lb1}) and (\ref{lb2}). So we see that the flavored ABJM spin chain cannot be generated from our fused model. Besides, by the same argument, we find the flavored ABJM spin chain cannot be obtained from another integrable boundary model $H_b$ (\ref{bm1}) either. So we guess that the flavored ABJM open spin chain is not integrable, though its integrability has been argued to hold from coordinate Bethe ansatz method.

\section{Conclusion and discussion}
In this paper we constructed the fused model of ABJM alternating spin chain by gluing two adjacent quantum spaces as well as two original auxiliary spaces. For closed spin chain, we proved the integrability of the fused model by constructing the R-matrix and showing the Yang-Baxter relation holds. We obtained the boost operator for the fused model based on the regularity condition of the fused R-matrix.  We also argued that the usual matrix type boost operator with local densities cannot exist in either of two original ABJM spin chains. For open spin chain, we calculated the concrete Hamiltonian for general fused K-matrices satisfying the reflection and dual reflection equations. We then analyzed the boundary terms of the Hamiltonian and found out some common structures of the component indices which are independent of the concrete choices of K-matrices. By comparing the boundary terms, we claimed that the previously studied flavored ABJM spin chain Hamiltonian cannot originate from the fused model or simply the combination of two original sub-chains Hamiltonians, and thus is expected to be non-integrable from the viewpoint of algebraic Bethe ansatz method.

There are several interesting directions for future research. Firstly, as we have mentioned in the main text, the ABJM spin chain can be seen as a three-site interacting model with homogeneous local Hamiltonian density $h_{j,j+1,j+2}$, in the sense that $h_{j,j+1,j+2}$ has the same expression no matter the starting site $j$ is odd or even. However, to construct the transfer matrix of the fused model, we have used two different Lax operators, $\mathcal{L}_{0\bar{0},2j-1}(u)$ on odd quantum site and $\mathcal{L}_{0\bar{0},2j}(u)$ on even quantum site. This remarkable difference makes us to search for a new construction of ABJM spin chain using one single Lax operator, say $L_{a,b,c}(u,\xi)$ with possibly additional inhomogeneity parameter $\xi$.

Secondly, we can continue to study the boost operator in ABJM spin chain. Though we have excluded the existence of matrix-type boost operators in the original two spin chains $\tau(u)$ and $\bar{\tau}(u)$, there are still possibilities to have other types of boost operators, such as differential operator which occurs in one-dimensional Hubbard model \cite{Links:2001ctb}, though the corresponding R-matrix is of non-difference form.  As for the fused model, since we have established the boost operator, the conserved charges can be related by the recursive relation $Q_{n+1}\sim \left[B,Q_n\right]$, and thus each charge will have a definite parity under spatial reflection transformation. Then we can investigate the so-called integrable initial state $|\Psi\rangle$ introduced in \cite{Piroli:2017sei} for ABJM fused spin chain, which is defined as a state annihilated by all the conserved charges with odd parities, i.e. $Q_{2n+1}|\Psi\rangle=0$. While for the original spin chain $\tau(u)$ or $\bar{\tau}(u)$, due to the lack of boost operator, there is no well-defined integrable state.

Finally, we could also consider some general problems addressed in the past studies of integrable models for our fused model. For instance, we can study the long range deformation of the fused model, where the boost of the conserved charge serves as an integrable deformation operator. Another challenging problem is to find a systematic way to classify the integrable alternating spin chain models, including the ABJM spin chain as a prototype.

\begin{appendix}
\end{appendix}

\end{document}